\documentclass[pdftex]{pasj01}
\usepackage{url}
\usepackage{color}

\Received{$\langle$reception date$\rangle$}
\Accepted{$\langle$acception date$\rangle$}
\Published{$\langle$publication date$\rangle$}

\begin{document}

\title{Suzaku X-ray observations of the mixed-morphology supernova remnant CTB~1}
\author{
Miho Katsuragawa\altaffilmark{1,2,}$^{*}$, 
Shinya Nakashima\altaffilmark{3}, 
Hideaki Matsumura\altaffilmark{2}, 
Takaaki Tanaka\altaffilmark{4}, 
Hiroyuki Uchida\altaffilmark{4}, 
Shiu-Hang Lee\altaffilmark{5},
Yasunobu Uchiyama\altaffilmark{6}, 
Masanori Arakawa\altaffilmark{6,7}, 
and Tadayuki Takahashi\altaffilmark{2,1}
}%
\altaffiltext{1}{Department of Physics, The University of Tokyo, Hongo, Bunkyo-ku, Tokyo 113-0033, Japan }
\altaffiltext{2}{Kavli Institute for the Physics and Mathematics of the Universe (WPI),The University of Tokyo Institutes for Advanced Study, The University of Tokyo, Kashiwa, Chiba 277-8583, Japan}
\altaffiltext{3}{RIKEN High Energy Astrophysics Laboratory, Hirosawa, Wako, Saitama 351-0198, Japan }
\altaffiltext{4}{Department of Physics, Kyoto University, Kitashirakawa Oiwake-cho, Sakyo-ku, Kyoto-shi, Kyoto 606-8502, Japan}
\altaffiltext{5}{Department of Astronomy, Kyoto University, Kitashirakawa Oiwake-cho, Sakyo-ku, Kyoto-shi, Kyoto 606-8502, Japan}
\altaffiltext{6}{Department of Physics, Rikkyo University, Nishi-Ikebukuro, Toshima-ku, Tokyo 171-8501, Japan}
\altaffiltext{7}{Astrophysical Big Bang Laboratory, RIKEN, Hirosawa, Wako, Saitama 351-0198, Japan}
\email{miho.katsuragawa@ipmu.jp}


\maketitle

\begin{abstract}
We present an X-ray study of the mixed-morphology supernova remnant CTB~1 (G116.9+0.2) observed with Suzaku.
The $0.6\mathchar`-\mathchar`-2.0~\mathrm{keV}$ spectra in the northeast breakout region of CTB~1 are well represented by a collisional ionization-equilibrium plasma model with an electron temperature of $\sim 0.3~\mathrm{keV}$, 
whereas those in the southwest inner-shell region can be reproduced by a recombining plasma model with an electron temperature of $\sim 0.2~\mathrm{keV}$, an initial ionization temperature of $\sim 3~\mathrm{keV}$, and an ionization parameter of $\sim 9 \times 10^{11}~\mathrm{cm^{-3}~s}$.
This is the first detection of the recombining plasma in CTB~1.
The electron temperature in the inner-shell region decreases outwards, which implies that the recombining plasma is likely formed by the thermal conduction via interaction with the surrounding cold interstellar medium.
The Ne abundance is almost uniform in the observed regions whereas Fe is more abundant toward the southwest of the remnant, suggesting an asymmetric ejecta distribution. 
We also detect a hard tail above the $2~\mathrm{keV}$ band that is fitted with a power-law function with a photon index of $2\mathchar`-\mathchar`-3$.
The flux of the hard tail in the $2\mathchar`-\mathchar`-10~\mathrm{keV}$ band is $\sim 5 \times 10^{-13}~\mathrm{erg~cm^{-2}~s^{-1}}$ and is peaked at the center of CTB~1.
Its origin is unclear but one possibility is a putative pulsar wind nebula associated with CTB~1.
\end{abstract}

\section{Introduction}
Mixed-morphology supernova remnants (MM SNRs) have a characteristic morphology of a radio shell with centrally-filled thermal X-rays \citep{rho1998}.
Most of the MM SNRs are likely interacting with a dense interstellar medium (ISM) as indicated by detections of CO emissions or OH masers at 1720~MHz in some cases (e.g., \cite{tatematsu1990}; \cite{rho1998}; \cite{yusef2003}). 
Some of them are also associated with GeV gamma-ray emissions (e.g., \cite{acero2016}), supporting the presence of a dense ISM in the vicinity of the remnants if the origin of the gamma-ray is ${\rm \pi}^{0}$ decays.
Such a dense surrounding environment probably plays an important role in the formation of MM SNRs, but detailed physical processes are still under debate (e.g., \cite{white1991}; \cite{cox1999}).

Recently, X-ray observations found recombining plasmas (RPs) in many MM SNRs (e.g., \cite{kawasaki2002}; \cite{yamaguchi2009}; \cite{ozawa2009}; \cite{ohnishi2011}; \cite{sawadakoyama2012}; \cite{uchida2012}; \cite{matsumura2017a}). 
An RP indicates the unusual evolution of thermal plasmas, because the ionization timescale is longer ($\gtrsim 10^5$~yr) than the typical age of SNRs in the typical density of electrons (0.1--1~cm$^{-3}$) and an ionizing plasma is expected in SNRs.
For the formation of RPs in MM SNRs, two scenarios are mainly considered.
One of the proposed scenarios is rapid decreasing of the electron temperature ($kT_e$) via the rarefaction of the plasma when the shock breaks out of the dense circumstellar medium into low density ISM \citep{itohmasai1989}. 
The other is the thermal conduction scenario, in which $kT_e$ decreases rapidly due to the interaction between the surrounding cold ISM and the hot plasmas (\cite{kawasaki2002}; \cite{matsumura2017a}).

CTB~1 (G116.9+0.2) is one of the MM SNRs.
It has a characteristic  morphology as shown in figure \ref{fig:image}; its non-thermal radio shell breaks at the northeast quadrant (\cite{landecker1982}; \cite{kothes2006}) and its thermal X-rays extend outward from the break region \citep{craig1997}.
Optical emission lines also show an incomplete shell morphology similar to the radio shell \citep{fesen1997}.
\citet{YarUyaniker2004} argued that CTB~1 is located inside a large H\emissiontype{I} bubble, and that the radio shell is interacting with the edge of the bubble.
No detections of gamma-ray emissions have been reported so far (e.g., \cite{acero2016}).
The distance to the remnant is highly uncertain in the range of 1.6--3.1~kpc (\cite{craig1997}; \cite{YarUyaniker2004}).
The SNR age is estimated to be a few $10^4$ years (\cite{KandH1991}; \cite{HandC1994}).

The nature of the thermal X-rays of CTB~1 has been studied with ASCA and Chandra observations by \citet{lazendic2006} and \citet{pannuti2010}.
Their results show that a primary component is a thermal plasma in collisional ionization equilibrium (CIE) with $kT_e \sim 0.2$--0.3~keV. 
However, they reported different minor components probably due to different analysis methods. 
\citet{lazendic2006} argued that another high-temperature plasma ($kT_e \sim 0.8$~keV) is present inside the radio shell, whereas \citet{pannuti2010} claimed a detection of a significant hard tail represented by a very high-temperature plasma ($kT_e \sim 3$~keV) or a power-law function with a photon index ($\Gamma \sim 2$--3) in both the inner radio shell and the breakout regions.
Therefore, the nature of the X-ray plasmas is still unclear. 

Here, we present new X-ray observations of CTB~1 with Suzaku \citep{mitsuda2007}.
Due to low instrumental background, the X-ray CCDs aboard Suzaku have very high sensitivity to diffuse sources in the 0.6--5.0~keV band and are suitable to reveal a precise physical property of CTB~1.  
Throughout this paper, statistical errors are quoted at a 68\% (1$\sigma$) confidence level.

\section{Observation and Data Reduction}
We observed the southwest (SW) and northeast (NW) regions of CTB~1 with the X-ray Imaging Spectrometer (XIS: \cite{koyama2007}) aboard the Suzaku satellite.
The XIS has three front-illuminated (FI) CCDs (XIS\,0, 2, 3) and one back-illuminated (BI) CCD (XIS\,1). 
The observation log is shown in table \ref{tab:observation_log}.
The total effective exposure time is $\sim82$~ks.

In order to evaluate a background spectrum, we also used nearby archival observations of GALACTIC PLANE 111 and ANTICENTER2 (hereafter BG1 and BG2, respectively).
As shown in table \ref{tab:observation_log}, the pointing direction of BG1 and BG2 are $\sim5\arcdeg$ away from CTB~1 in the Galactic latitude, but are on the Galactic plane.
Therefore, an average spectrum of BG1 and BG2 likely represents a background spectrum at the coordinates of CTB~1. 
We also confirmed that no bright sources are included in the XIS field-of-views (FoVs) of BG1 and BG2.

We processed the above XIS data with the HEADAS software version 6.20 and the calibration database released in December 2016.
We started with cleaned event lists produced by the standard pipeline process.
The data taken by the 3$\times$3 and 5$\times$5 editing modes were merged.
We discarded cumulative flickering pixels and pixels adjacent to the flickering pixels according to the noisy pixel maps provided by the XIS team\footnote{\url{http://www.astro.isas.jaxa.jp/suzaku/analysis/xis/nxb_new2}}.
Events below 0.6 keV were not used because contamination of O  \emissiontype{I} K$\alpha$ from the sunlit-Earth atmosphere in our observations was more severe than that shown in  \citet{sekiya2014}, and spectral modeling below 0.6 keV is highly uncertain even when we add a Gaussian for O  \emissiontype{I} K$\alpha$.
The XIS\,2 and a part of the XIS\,0 have not been functional since anomalies occurred in 2006 November and in 2009 June, respectively. 
We therefore do not used the entire XIS\,2 and segment A of XIS\,0. 

\begin{table*}[t]
	\caption{Observation log.}
	\centering
		\begin{tabular}{lcccccccccccc}\hline
			\multicolumn{1}{c}{Target Name} & Ods. ID & Obs. Date & $l$ & $b$ & Effective Exposure \\ \hline \hline
			CTB~1\_SW & 506034010 & 2011-12-29 & 116\fdg8913 & 0\fdg2952 & 29 ks\\
			CTB~1\_NE & 506035010 & 2011-12-28 & 117\fdg1835 & 0\fdg1796 & 53 ks \\
			GALACTIC PLANE 111 (BG1) & 501100010 & 2006-06-06 & 111\fdg5011 & 1\fdg3149 & 72 ks\\
			ANTICENTER2 (BG2) & 503006010 & 2008-08-01 & 122\fdg9896 & 0\fdg0395 & 86 ks\\ \hline
		\end{tabular}
	\label{tab:observation_log}
\end{table*}

\section{Analysis}

\subsection{XIS Images of CTB~1}
\begin{figure}[!t]
	\begin{center}
		\includegraphics[width=75mm]{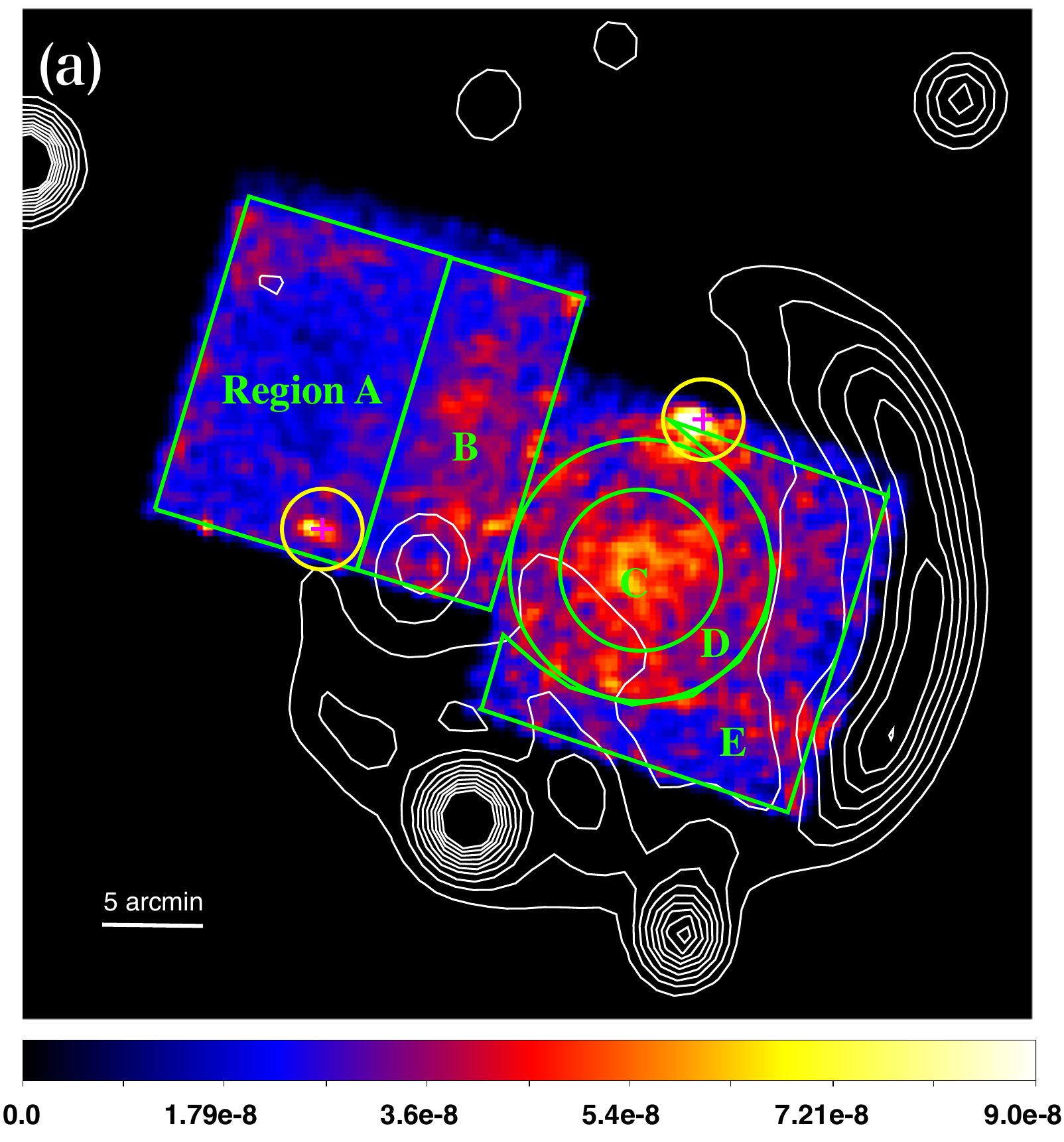}
		\includegraphics[width=75mm]{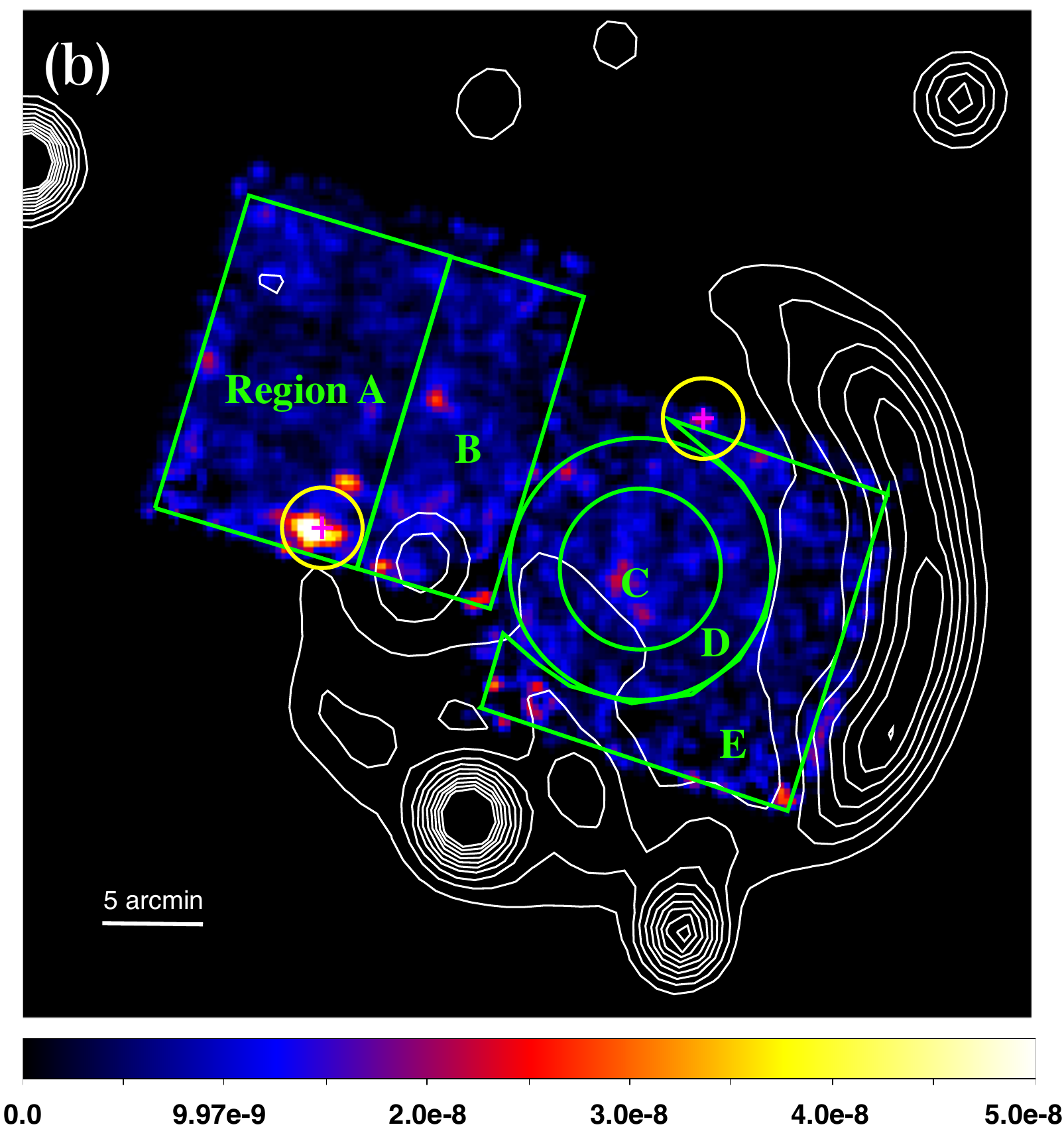}
	\end{center}
	\caption{Exposure-corrected XIS images of CTB~1 in the (a) 0.6--2.0 keV band and (b) 2.0--5.0 keV band.
	The contours are a radio intensity map observed with the Canadian Galactic Plane Survey at 408 MHz \citep{kothes2006}.  
	The magenta crosses with the yellow circles indicates bright point sources excluded in the spectral analyses.
	The green regions labeled with A, B, C, D, and E are spectral extraction regions (see section \ref{secSRA}). 
	}
  \label{fig:image}
\end{figure}

Figure \ref{fig:image} shows exposure-corrected XIS images of CTB~1 in the 0.6--2.0~keV band and the 2.0--5.0~keV band.
The data of XIS\,0, 1, and 3 were co-added, and the non X-ray backgrounds (NXBs) estimated by {\tt xisnxbgen} \citep{tawa2008} were subtracted.
As already reported by \citet{lazendic2006} and \citet{pannuti2010}, soft X-ray emission in the 0.6--2.0~keV band is concentrated at the center of the radio shell and extends toward the break of the radio shell.
On the other hand, no remarkable diffuse structures are seen in the 2.0--5.0 keV band image.

We found two bright point sources at the southern edge of the NE observation and the northern edge of the SW observation.
They were also identified in the 2XMMi catalog \citep{watson2009} and have fluxes of $\sim 4\times10^{-13}$~erg~cm$^{-2}$~s$^{-1}$.
In the following spectral analyses, we excluded those point sources by the circular regions with $2\arcmin$ radii as shown in figure \ref{fig:image}.

\subsection{Spectral modeling}
In this section, we performed spectral modeling using the XSPEC version 12.9.1a.
We extracted 0.6--10.0 keV and 0.6--8.0 keV spectra from the FI CCDs and BI CCD, respectively, and then subtracted corresponding NXB spectra created by {\tt xisnxbgen} from them.
Redistribution matrix files and auxiliary response files were generated by {\tt xisrmfgen} and {\tt xissimarfgen} \citep{ishisaki2007}, respectively.
The FI and BI CCDs spectra were simultaneously fitted.

\subsubsection{Background Estimation}
\label{secBGE}
\begin{figure}[t]
  \begin{center}
    \includegraphics[width=80mm]{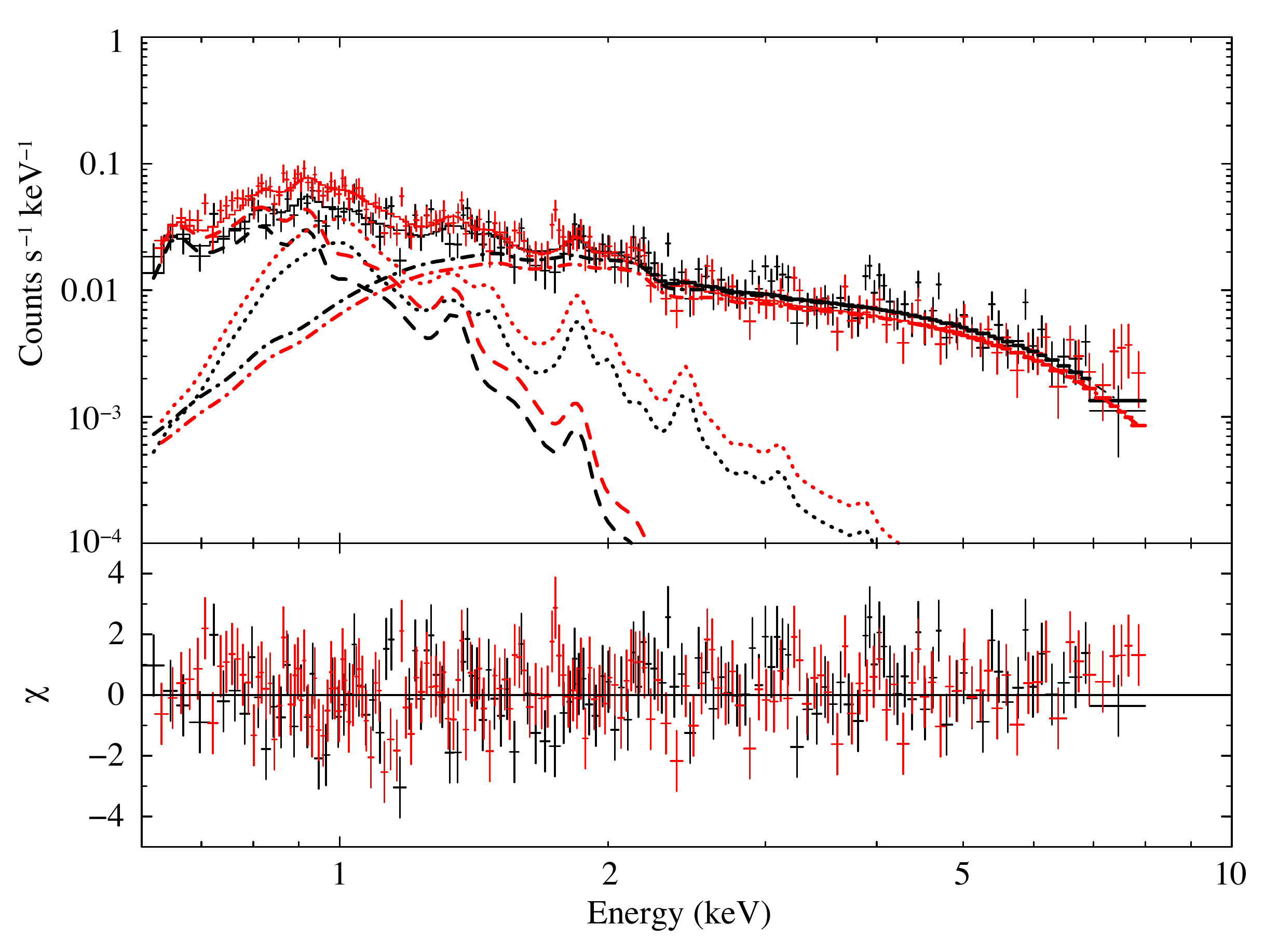}
  \end{center}
  \caption{XIS\,1 spectra of BG1 (black) and BG2 (red) after the NXB subtraction.
The solid curves show the Galactic background model described in section \ref{secBGE}.
The dashed, dotted, and dash-dotted curves indicate the APEC1, the APEC2, and the CXB components, respectively.
  }
  \label{fig:ganda_model}
\end{figure}

Before analyzing the spectra of CTB~1, we estimated background spectra using the BG1 and BG2 observations. 
Figure \ref{fig:ganda_model} shows spectra extracted from the entire FoVs of those background observations.
Continua dominate the spectra and weak emission lines were seen below 2~keV.

The X-ray background of the anti-Galactic center direction likely consists of a superposition of unresolved stellar sources and the cosmic X-ray background (CXB) (e.g., \cite{masui2009}).
The former component was approximated by two thin thermal plasma models in CIE, because individual stellar sources show a broad range of average temperatures \citep{gudel2009}.
We used the APEC code with the AtomDB version 3.0.8 to reproduce the plasma spectra \citep{foster2012}.
The abundance ($Z$) were fixed to the solar value of \citet{AandG1989} while the temperatures and the normalizations ($N$) were treated as free parameters.   
Those plasma models are subject to the absorption due to the foreground cold ISM, which was modeled by the PHABS code in XSPEC (hereafter PHABS1).
The absorption column density ($N_\mathrm{H}$) of PHABS1 was allowed to vary.
The CXB component was modeled by a power-law function (PL) with $\Gamma$ of 1.4 and the 2--10~keV surface brightness of $6.9 \times 10^{-15}$~erg~cm$^{-2}$~s$^{-1}$~arcmin$^{-2}$ according to \citet{kushino2002}.
The CXB component is subject to the Galactic absorption (PHABS2) of which column density was estimated from the {\tt nh} tool \citep{kalberla2005}.  
In summary, our model is expressed as 
\begin{eqnarray}
  \mathrm{CONST \times PHABS1 \times (APEC1 + APEC2)} \nonumber \\
  \mathrm{+ PHABS2 \times PL}, \nonumber
\end{eqnarray}
where CONST represents the difference in the surface brightness of the stellar component between BG1 and BG2.
CONST of BG1 was fixed to 1.0 while that of BG2 was allowed to vary.
The BG1 and BG2 spectra were simultaneity fitted with the above model. 
The model parameters are common between BG1 and BG2 except for CONST and PHABS2.

The best-fit model and parameters are shown in figure \ref{fig:ganda_model} and table \ref{tab:bg_best_fit_parameter}, respectively.
The model represents the spectra with $\chi^2_\nu = 1.21$.
The obtained plasma temperatures of 0.21 keV and 0.96 keV agree with the typical temperature range of coronal X-rays \citep{JandG2015}.
We found that the surface brightness of the stellar component is different by a factor of 2 between BG1 and BG2.
Since CTB~1 is located in the middle of BG1 and BG2, CONST was fixed at the average of them (1.5) in following spectral analyses of CTB~1.
We confirmed that our results were not affected even when CONST was fixed to 1 or 2.

\begin{table}[t]
\tbl{Best-fit parameters of the background\footnotemark[$*$].}
{
\begin{tabular}{lll}
	\hline
	Model & Parameter (unit) & Value \\ 
	\hline
	CONST & factor of BG1 & 1.0 (fixed) \\
	             & factor of BG2 & 2.0$\pm$0.1  \\ 
	\hline
	PHABS1 & $N_\mathrm{H}$ ($10^{21}$ cm$^{-2}$) & 5.8$\pm$0.5 \\ 
	\hline
	APEC1 & $kT_e$ (keV) & 0.21$^{+0.01}_{-0.02}$  \\
		     & $Z$ (solar) & 1.0 (fixed)  \\
	             & $N$\footnotemark[$\dagger$] ($10^{11}$ cm$^{-5}$) & 1.6$^{+0.9}_{-0.4}$   \\ 
	\hline
	APEC2 & $kT_e$ (keV) & 0.96$^{+0.05}_{-0.04}$   \\
		    & $Z$ (solar) & 1.0 (fixed)   \\
		    & $N$\footnotemark[$\dagger$]  ($10^{11}$ cm$^{-5}$)  & 0.16$\pm$0.02  \\ 
	\hline
	PHABS2 & $N_\mathrm{H}$ of BG1 ($10^{21}$ cm$^{-2}$) & 7.63 (fixed) \\
		      & $N_\mathrm{H}$ of BG2 ($10^{21}$ cm$^{-2}$) & 7.38 (fixed) \\ 
	\hline
	PL & $\Gamma$ & 1.4 (fixed)  \\
	     & surface brightness\footnotemark[$\ddagger$]  & 5.3 (fixed) \\ 
	\hline
	$\chi^2$/d.o.f. & & 875.52/726  \\ \hline \vspace{-3mm} \\
\end{tabular}
}
\label{tab:bg_best_fit_parameter}
\begin{tabnote}
\hangindent6pt\noindent
\hbox to6pt{\footnotemark[$*$]\hss}\unskip%
Parameters are common between BG1 and BG2 except for CONST and PHABS2.
\par
\hangindent6pt\noindent
\hbox to6pt{\footnotemark[$\dagger$]\hss}\unskip%
Normalization of the APEC component is $\frac{1}{4\pi D^2} \int n_en_\mathrm{H}dV$, where $D$ is the distance to the source, $n_e$ and $n_\mathrm{H}$ are the electron and hydrogen densities, respectively, and $V$ is the X-ray emitting volume.
\par
\hangindent6pt\noindent
\hbox to6pt{\footnotemark[$\ddagger$]\hss}\unskip%
Unit of 10$^{-15}$ erg~cm$^{-2}$~s$^{-1}$~arcmin$^{-2}$ in the 2.0--10.0~keV band.
\end{tabnote}
\end{table}

\subsubsection{Spectra of the NE region}
\begin{figure}[t]
	\begin{center}
	\includegraphics[width=80mm]{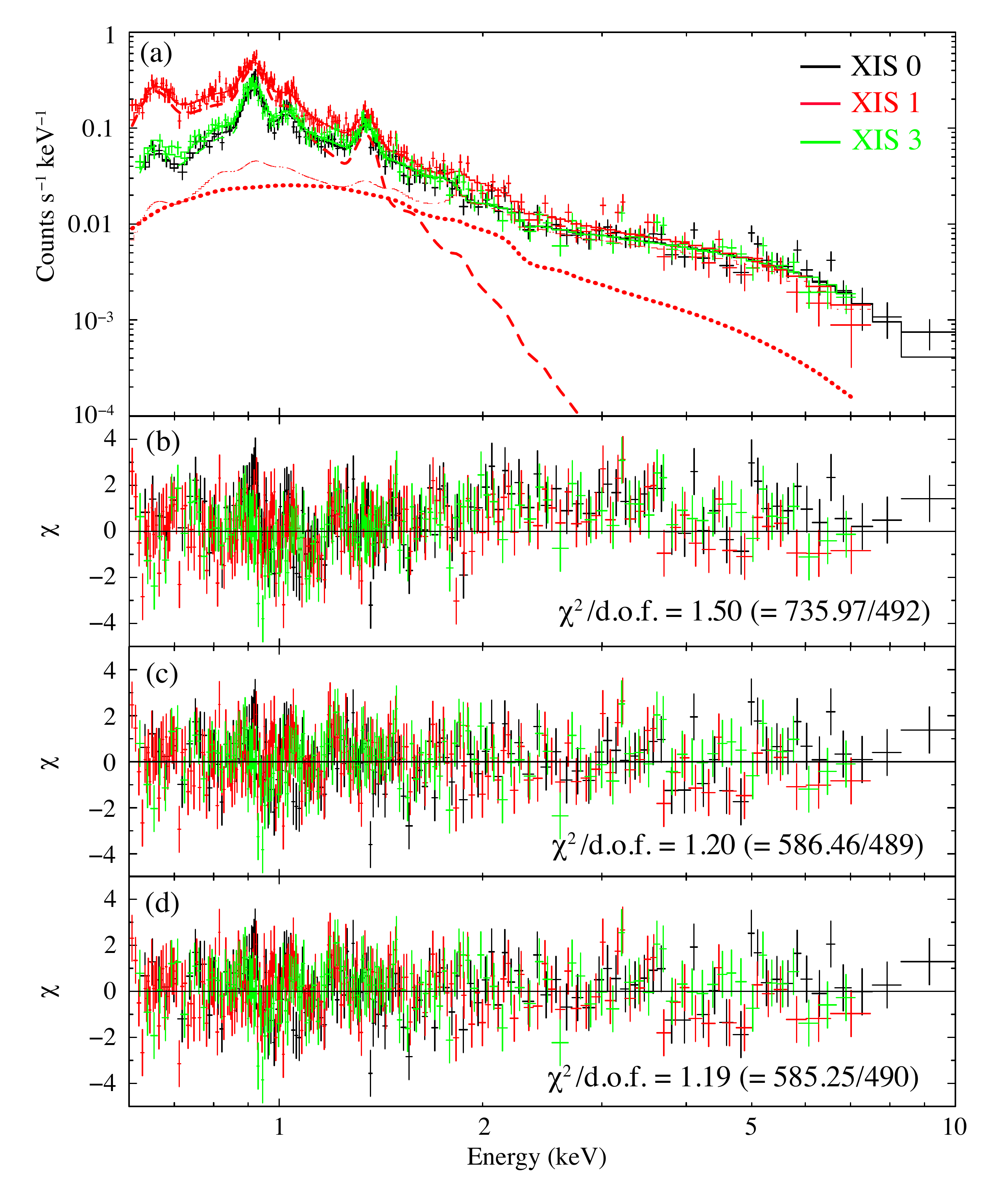}
	\end{center}
	\caption{Top panel (a) shows spectra of the NE observation taken by the XIS\,0 (black), 1 (red) and 3 (green). 
	The solid curves are the NE Model3. 
	The dashed, dotted, and dash-dotted curves represent the VAPEC, the PL, and the background components of XIS\,1, respectively.
	The lower three panels show the residuals of the data compared to the NE Model1(b), the NE Model2 (c) and  the NE Model3 (d).
	}
	\label{fig:ctb1_ne}
\end{figure}

Spectra extracted from the entire FoV of the NE observation of CTB~1 are shown in figure \ref{fig:ctb1_ne}(a).
Strong emission lines of highly ionized O, Ne, and Mg present below 2~keV while no remarkable features are seen above 2~keV. 

According to \citet{pannuti2010}, we modeled those spectra by a single CIE plasma model with variable abundances of individual metals (the VAPEC model).
Abundances of Ne, Mg, Si, and Fe were allowed to vary.
Abundances of S and Ni were linked to those of Si and Fe, respectively.
If the O abundance was treated as a free parameter, it was not constrained.
That is because the intensity of the radiative recombination continua from O is much higher than that of the bremsstrahlung from H for a plasma with $kT_e =$ 0.2--0.3~keV. 
We therefore fixed the O abundance to the solar value.
Other metals were fixed to the solar abundances.
Absorption of the ISM was represented by the PHABS model.
We also included the background model constructed in the previous section.
In the background model, $N_\mathrm{H}$ of PHABS2 was fixed to $5.96\times10^{21}$ cm$^{-2}$ estimated by the {\tt nh} tool.
Hereafter, we refer to this model as NE Model1.

The best-fit parameters are shown in table \ref{tab:ctb1ne_best_fit_parameter} and the residuals between the data and the NE Model1 are shown in figure \ref{fig:ctb1_ne}(b).
The model almost reproduces the observed spectra below 2~keV, but we found an excess of the data above 2~keV with $\chi^2_\nu=1.50$.
It suggests that an additional hard component is necessary as reported by \citet{pannuti2010}.
We therefore tried two different models; one includes another APEC model (NE Model2) and the other includes a power-law function (NE Model3).  

\begin{table*}[t]
	\tbl{Best-fit parameters of the NE spectra.}
	{
	\begin{tabular}{lllll}
		\hline
		Model               & Parameter (unit)                     & NE Model1                          & NE Model2                           & NE Model3                        \\
		\hline
		PHABS             & $N_\mathrm{H}$ ($10^{21}$cm$^{-2}$) & 3.5$\pm$0.2      & 2.4$^{+0.2}_{-0.3}$              & 2.1$^{+0.3}_{-0.4}$            \\
		\hline
		VAPEC             & $kT_e$ (keV)                          & 0.293$^{+0.005}_{-0.004}$ & 0.295$^{+0.004}_{-0.003}$  & 0.30$\pm$0.05                    \\
		                         & $Z_\mathrm{Ne}$ (solar)        & 1.30$\pm$0.06                    & 1.9$^{+0.15}_{-0.08}$          & 2.2$^{+0.3}_{-0.2}$             \\
			                 & $Z_\mathrm{Mg}$ (solar)       & 0.98$\pm$0.09                    & 1.6$\pm$0.1                        & 1.9$^{+0.4}_{-0.3}$             \\
			                 & $Z_\mathrm{Si,\ S}$ (solar)   & 0.9$\pm$0.2                         & $<$ 1.3                                & $<$1.6                           \\
			                 & $Z_\mathrm{Fe,\ Ni}$ (solar) & 0.14$\pm$0.02                     &0.22$\pm$0.02                     & 0.23$\pm$0.03                    \\
			                 & $Z_\mathrm{Others}$ (solar) & 1.0\ (fixed)                            & 1.0\ (fixed)                           & 1.0\ (fixed)                              \\
					& $N$\footnotemark[$*$] ($10^{12}$ cm$^{-5}$) & 0.96$\pm$0.08 & 0.5$^{+0.07}_{-0.08}$            & 0.40$^{+0.10}_{-0.09}$       \\
		\hline
		APEC               & $kT_e$ (keV)                         & -                                            & 2.2$^{+0.2}_{-0.3}$             & -                                           \\
		                         & $Z$ (solar)                             & -                                            & $<$ 0.031                             & -                                           \\
		                         & $N$\footnotemark[$*$] ($10^{12}$ cm$^{-5}$) & -                 & 0.097$^{+0.013}_{-0.008}$ & -                                           \\
		\hline
		PL                     & $\Gamma$                            & -                                            & -                                           & 2.7$\pm$0.2                        \\
		                         & surface brightness\footnotemark[$\dagger$]    & -                & -                                           & 1.2$\pm$0.2                    \\
		\hline
		$\chi^2$/d.o.f.  &                                               & 735.97/492                            & 586.46/489                          & 585.25/490                          \\
		\hline  \vspace{-3mm} \\
	\end{tabular}
	}
	\label{tab:ctb1ne_best_fit_parameter}
	\begin{tabnote}
		\hangindent6pt\noindent
		\hbox to6pt{\footnotemark[$*$]\hss}\unskip%
		Normalization of the APEC component is $\frac{1}{4\pi D^2} \int n_en_\mathrm{H}dV$, where $D$ is the distance to the source, $n_e$ and $n_\mathrm{H}$ are the electron and hydrogen densities, respectively, and $V$ is the X-ray emitting volume.
		\par
		\hangindent6pt\noindent
		\hbox to6pt{\footnotemark[$\dagger$]\hss}\unskip%
		Unit of 10$^{-15}$ erg~cm$^{-2}$~s$^{-1}$~arcmin$^{-2}$ in the 2.0--10.0~keV band.
	\end{tabnote}
\end{table*}	

In the NE Model2, the temperature, the normalization, and the abundance relative to the solar values in the APEC were allowed to vary.
The resultant best-fit parameters and residuals are shown in table \ref{tab:ctb1ne_best_fit_parameter} and figure \ref{fig:ctb1_ne}(c), respectively.
The fit statistic was remarkably reduced to $\chi^2_\nu=1.20$.
The parameters of the VAPEC component in the NE Model2 was almost consistent with those in the NE Model1.
The APEC component shows $kT_e$ of 2.2~keV and the abundance of $< 0.031$~solar.
Such a high temperature plasma has been observed only in young supernova remnants.
Moreover, the very low abundance does not agree with the abundance of the ejecta nor the ISM.  
We thus consider that the NE Model2 is physically unreasonable.

The best-fit parameters and residuals of the NE Model3 are shown in table \ref{tab:ctb1ne_best_fit_parameter} and figure \ref{fig:ctb1_ne}(d), respectively.
This model gave the lowest $\chi^2_\nu$ of 1.19.
The absorption column density and the metal abundances were slightly changed from the NE Model1.
The photon index $\Gamma$ of the PL component is 2.7, and the surface brightness in the 2--10~keV band is lower than that of the CXB component by a factor of $\sim5$.
We consider that the NE Model3 is the best representative of the spectra.

\subsubsection{Spectra of the SW region}
\begin{figure}[t]
	\begin{center}
	\includegraphics[width=80mm]{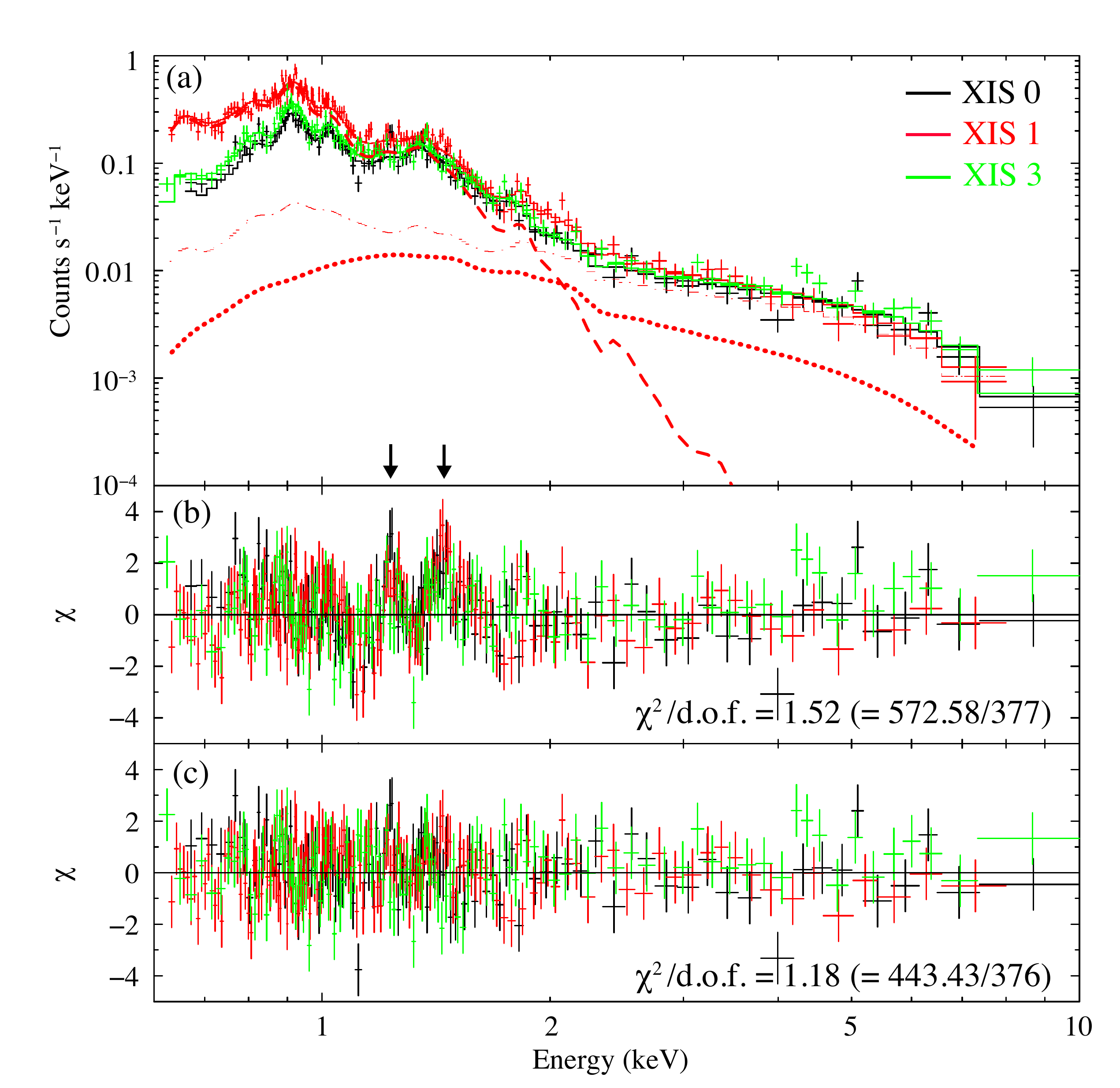}
	\end{center}
	\caption{
	Top panel (a) shows spectra of the SW observation taken by the XIS\,0 (black), 1 (red) and 3 (green). 
	The solid curves are the SW Model2. 
	The dashed, dotted, and dash-dotted curves represent the VRNEI, the PL, and the background components of XIS\,1, respectively. 
	The lower two panels show the residuals of the data compared to the SW Model1 (b), the SW Model2 (c).
	The two arrows point at 1.23~keV and 1.45~keV, respectively.
	}
	\label{fig:ctb1_sw}
\end{figure}

Spectra extracted from the entire FoV of the SW observation of CTB~1 are shown in figure \ref{fig:ctb1_sw}(a).
They are similar to the NE spectra, but have hump-like structures in the 1.1--1.7~keV band. 

We first fitted the SW spectra with the same model as the NE Model3 (hereafter SW Model1).
The absorption column density of the CXB component estimated from the {\tt nh} tool is $6.35 \times 10^{21}$~cm$^{-2}$.
The best-fit parameters and the residuals are shown in table \ref{tab:ctb1sw_best_fit_parameter} and figure \ref{fig:ctb1_sw}(b), respectively.
The fit statistic is $\chi^2_\nu=1.52$, and sawtooth like residuals are shown around $\sim 1.23$~keV and $\sim 1.45$~keV.
These energies correspond to the centroid energies of Fe L lines (e.g., Fe \emissiontype{XXI} 1s2 2s1 2p2 4s1$\rightarrow$1s2 2s1 2p3 and Fe\emissiontype{XXI} 1s2 2s2 2p1 5d1$\rightarrow$1s2 2s2 2p2), but these emissions are not strong in a plasma with $kT_e$ of $\sim 0.3$~keV.
In order to improve the residuals, we tried to add another plasma component with a high $kT_e$ because these Fe L lines have a peak at $kT_e \sim 0.9$~keV, but a suitable model could not be found.

\begin{figure}[t]
	\begin{center}
	\includegraphics[width=75mm]{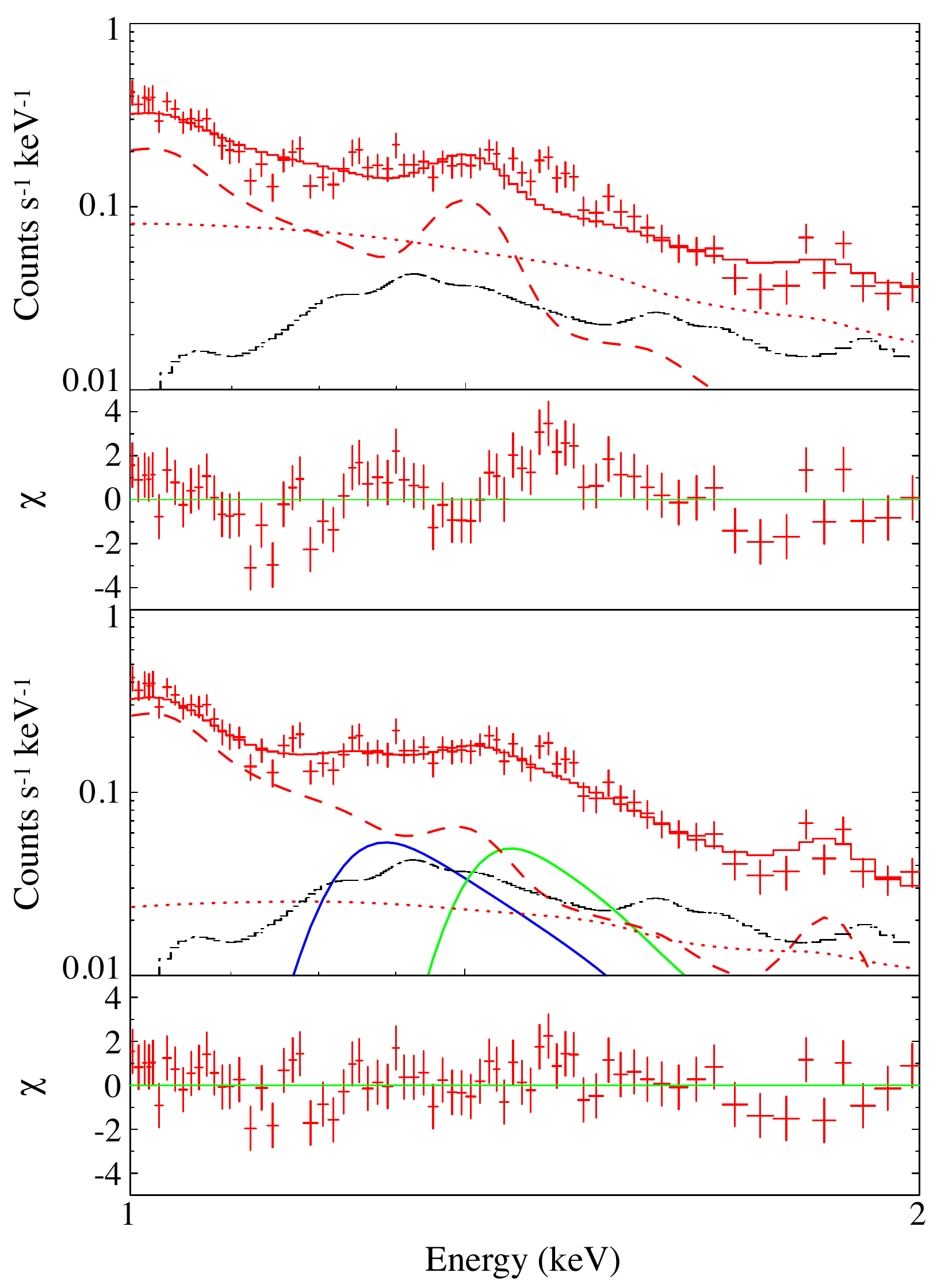}
	\end{center}
	\caption{
	Top and third panels show spectra (red crosses) of the SW observation taken by XIS\,1 with the best-fit models (red solid lines).
	In the top panel, the model consists of the VAPEC (red dashed), PL (red dotted) and background (black dashed).
	In the third panel, the model is same as that in the top panel but it additionally includes two REDGE components with the edge energies of 1.189~keV (blue solid) and 1.34~keV (green solid).
	Second and the lower panels show the residuals of the data compared to the models.
	}
	\label{fig:ctb1_sw_redge}
\end{figure}

Sawtooth like residuals are often found in MM SNRs with RPs (e.g., \cite{yamaguchi2009}) because of a lack of radiative recombination continua (RRCs).
The residuals around $\sim 1.23$~keV and $\sim 1.45$~keV indicate the excess of RRCs of He-like and H-like Ne, which are clear evidence of an RP.
To investigate this hypothesis, we added two REDGE models to the SW Model1. 
Figure \ref{fig:ctb1_sw_redge} shows the result of the spectra fitting.
We confirmed that the fit was significantly improved and the electron temperature of the REDGE models is 0.13$^{+0.04}_{-0.03}$~keV.
The best-fit energies of the REDGE models are 1.20$\pm$0.02~keV and 1.34$^{+0.04}_{-0.03}$~keV, which are consistent with the edge energies of He-like and H-like Ne (1.196~keV and 1.362~keV), respectively. 

In order to apply an RP model to the full-band spectra, we used the VRNEI model, which takes into account ion fractions and ionization timescales of all elements in the recombination-dominant state.
We therefore replace the VAPEC model to the VRNEI model and refer to this model as SW Model2.
The VRNEI model calculates the spectrum of a non-equilibrium ionization plasma after a rapid transition of the electron temperature from $kT_{\rm init}$ to $kT_e$ with a recombining timescale ($n_et$).
Since the initial temperature $kT_{\rm init}$ was not well constrained in our data, we fixed it to 3~keV, in which most Ne and Mg ions become bare nuclei.
The best-fit parameters and the residuals are shown in table \ref{tab:ctb1sw_best_fit_parameter} and figure \ref{fig:ctb1_sw}(c), respectively.
No remarkable residuals are seen in the SW Model2, and the fit statistic is significantly improved to $\chi^2_\nu=1.18$.
The obtained $n_et = 9.3\times10^{11}$~s~cm$^{-3}$ indicates that the plasma in the SW region is indeed an RP.

\begin{table}[t]
	\tbl{Best-fit parameters of the SW spectra.}
	{
	\begin{tabular}{llll}
		\hline
		Model      & Parameter (unit)                       & SW Model1                            & SW Model2                                  \\
		\hline
		PHABS    & $N_\mathrm{H}$ ($10^{21}$cm$^{-2}$) & 2.9$\pm$0.4             & 4.5$\pm$0.2                                \\
		\hline
		VAPEC     & $kT_e$ (keV)                           & 0.298$^{+0.008}_{-0.007}$   & 0.186$^{+0.005}_{-0.004}$          \\
		/VRNEI     & $kT_\mathrm{init}$  (keV)        &  -                                            & 3.0 (fixed)                                        \\
			         & $Z_\mathrm{Ne}$ (solar)         & 2.0$^{+0.4}_{-0.3}$                & 2.9$\pm$0.3                                \\
			         & $Z_\mathrm{Mg}$ (solar)         & 1.8$^{+0.5}_{-0.4}$                & 1.0$\pm$0.2                              \\
			         & $Z_\mathrm{Si,\ S}$ (solar)     & $<$2.6                                    & 0.4$\pm$0.2                                \\
			         & $Z_\mathrm{Fe,\ Ni}$ (solar)   & 0.43$^{+0.07}_{-0.06}$          & 2.1$^{+0.3}_{-0.2}$                     \\
			         & $Z_\mathrm{Others}$ (solar)   & 1.0\ (fixed)                                  & 1.0\ (fixed)                                       \\
			         & $N$\footnotemark[$*$] ($10^{12}$ cm$^{-5}$) & 0.6$^{+0.2}_{-0.1}$ & 3.0$^{+0.5}_{-0.4}$         \\
			         & $n_e t$ ($10^{11}$s cm$^{-3}$)  & -                                           & 9.3$\pm$0.4                               \\
		\hline
		PL             & $\Gamma$                                & 3.5$\pm$0.1                           & 2.2$\pm$0.3                               \\
			         & surface brightness\footnotemark[$\dagger$] & 1.6$^{+0.2}_{-0.1}$ & 1.7$^{+0.3}_{-0.2}$ \\
		\hline
		$\chi^2$/d.o.f.  &                                            & 572.58/377                             & 443.43/376                                 \\
		\hline \vspace{-3mm} \\
	\end{tabular}
	}
	\label{tab:ctb1sw_best_fit_parameter}
	\begin{tabnote}
		\hangindent6pt\noindent
		\hbox to6pt{\footnotemark[$*$]\hss}\unskip%
		Normalization of the APEC component is $\frac{1}{4\pi D^2} \int n_en_\mathrm{H}dV$, where $D$ is the distance to the source, $n_e$ and $n_\mathrm{H}$ are the electron and hydrogen densities, respectively, and $V$ is the X-ray emitting volume.
		\par
		\hangindent6pt\noindent
		\hbox to6pt{\footnotemark[$\dagger$]\hss}\unskip%
		Unit of 10$^{-15}$ erg~cm$^{-2}$~s$^{-1}$~arcmin$^{-2}$ in the 2.0--10.0~keV band.
	\end{tabnote}
\end{table}

\subsubsection{Spatially resolved spectra}
\label{secSRA}
We found that the plasma in the NE region is in the CIE state while that in the SW region is in the recombining phase.
In order to investigate more detailed spatial variation of the plasma properties, we divided the FoVs into five regions (A, B, C, D and E) as shown in figure \ref{fig:image} according to the 0.6--2.0~keV morphology.

\begin{figure*}[p]
	\begin{center}
	\includegraphics[width=165mm]{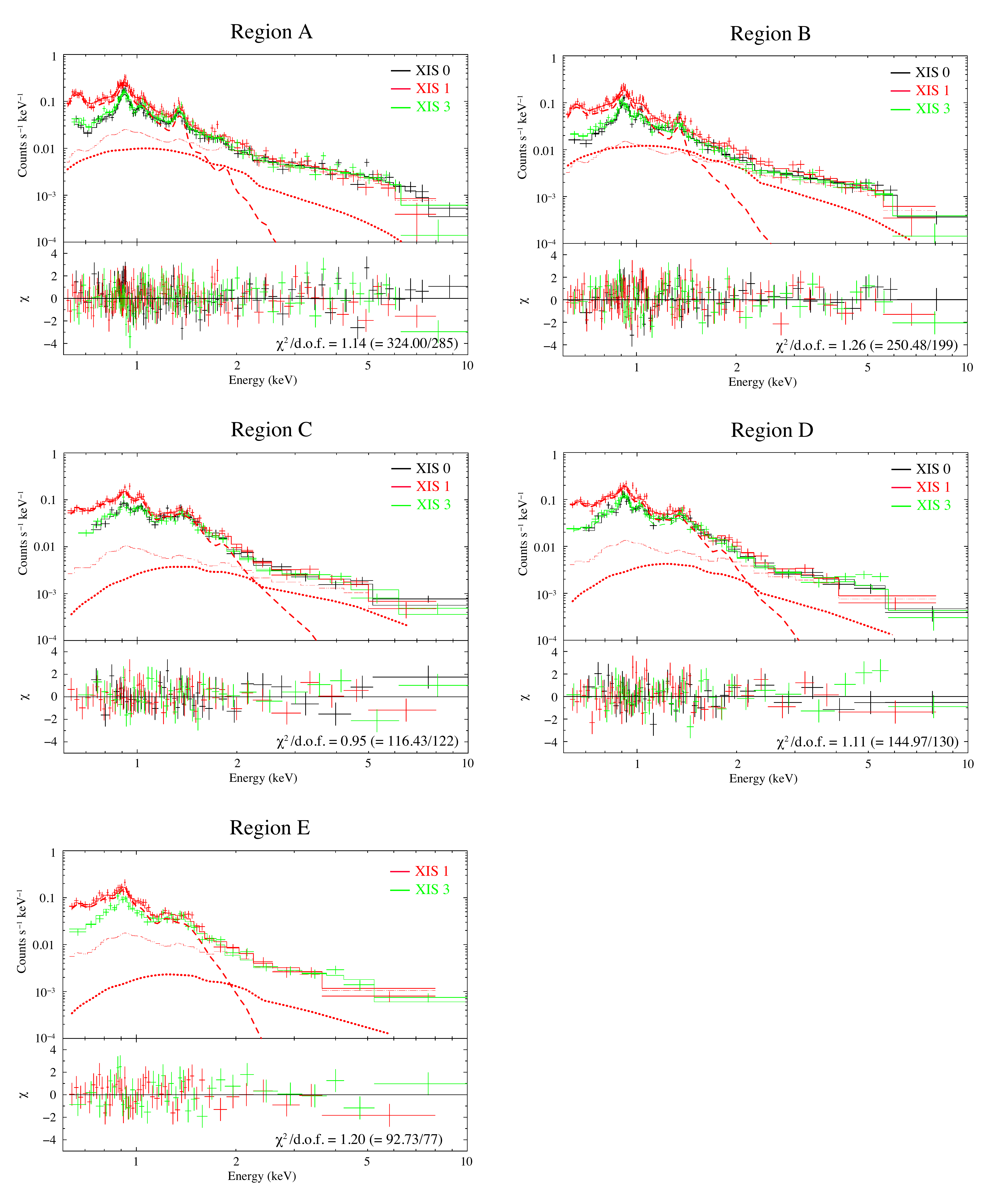}
	\end{center}
	\caption{
	Each panel shows spectra extracted from each region taken by the XIS\,0 (black), 1 (red) and 3 (green) and residuals from the best-fit models.
	The solid curves are the best-fit model. 
	The dashed, dotted, and dash-dotted curves represent the VAPEC (regions A and B)/VRNEI (regions C, D and E), the PL, and the background components of XIS\,1, respectively. 
	}
	\label{fig:ctb1_ABCDE}
\end{figure*}

The regions A and B are included in the NE observation, and therefore, we fitted their spectra with the NE Model3.  
On the other hand, we applied SW Model2 to spectra of the regions C, D, and E, which are included in the SW observation.
Since the non-functioning segment of XIS~0 occupies a large area in the region~E, we did not use the XIS~0 spectrum extracted from the region~E.
We show the best-fit parameters and the fitted spectra in table \ref{tab:ctb1ABCDE_best_fit_parameter} and figure \ref{fig:ctb1_ABCDE}, respectively.
All the spectra are well reproduced by the models. 
The photon index in the region E is fixed at 2.2, which is the best-fit value of the entire SW region, because that was not constrained in the fitting.
We found significant spatial variations in $N_\mathrm{H}$, $kT_e$, the metal abundances, and $n_et$.   

\begin{table*}[t]
	\tbl{Best-fit parameters of spectra for the spatial resolved analysis.}
	{
	\begin{tabular}{lllllll}
		\hline
		Model        & Parameter (unit)                          & region A                               & region B                               & region C                             & region D                    & region E                     \\
		\hline
		Absorption & $N_\mathrm{H}$ ($10^{21}$cm$^{-2}$) & 2.3$\pm$0.4            & 2.2$\pm$0.5                        & 4.4$^{+0.4}_{-0.5}$             & 4.4$\pm$0.4             & 4.3$^{+0.5}_{-0.6}$    \\
		\hline
		VAPEC      & $kT_e$ (keV)                             & 0.288$^{+0.007}_{-0.006}$  & 0.306$^{+0.010}_{-0.009}$  & 0.193$^{+0.010}_{-0.009}$ & 0.20$\pm$0.01         & 0.155$\pm$0.009       \\
		/VRNEI      & $kT_\mathrm{init}$ (keV)           & -                                            & -                                           & 3.0 (fixed)                            & 3.0 (fixed)                 & 3.0 (fixed)                   \\
			          &  $Z_\mathrm{Ne}$ (solar)          & 1.9$^{+0.3}_{-0.2}$              & 2.4$^{+0.5}_{-0.4}$              & 2.9$^{+0.6}_{-0.5}$             & 2.7$^{+0.5}_{-0.4}$  & 3.2$^{+0.9}_{-0.6}$     \\
			          &  $Z_\mathrm{Mg}$ (solar)          & 1.8$^{+0.5}_{-0.3}$              & 1.6$^{+0.5}_{-0.4}$              & 0.8$\pm$0.3                        & 1.3$^{+0.4}_{-0.3}$  & $<$1.7                        \\
			          &  $Z_\mathrm{Si,\ S}$ (solar)      & $<$2.7                                  & $<$2.9                                 & 0.4$\pm$0.2                        & 0.6$^{+0.4}_{-0.5}$   & $<$4.0                        \\
			          &  $Z_\mathrm{Fe,\ Ni}$ (solar)    & 0.20$\pm$0.03                     & 0.30$^{+0.06}_{-0.05}$        & 1.4$\pm$0.5                        & 1.2$^{+0.6}_{-0.4}$  & $>$2.4                         \\
			          &  $Z_\mathrm{Others}$ (solar)    & 1.0 (fixed)                             & 1.0 (fixed)                            & 1.0 (fixed)                            & 1.0 (fixed)                 & 1.0 (fixed)                    \\
			          & $N$\footnotemark[$*$] ($10^{12}$ cm$^{-5}$) & 0.5$\pm$0.1  & 0.4$^{+0.2}_{-0.1}$              & 3.3$\pm$0.8                        & 2.7$^{+0.8}_{-0.7}$  & 2.6$^{+1.1}_{-0.8}$     \\
			          & $n_e t$ ($10^{11}$s cm$^{-3}$) & -                                           & -                                            & 7.0$\pm$0.5                        & 9.9$^{+0.8}_{-0.6}$  & 10.5$\pm$1.0              \\
		\hline
		PL              & $\Gamma$                                  & 2.7$\pm$0.4                        & 2.6$\pm$0.2                         & 1.9$^{+0.5}_{-0.6}$             & 2.6$\pm$0.7             & 2.2 (fixed)                   \\
			         & surface brightness\footnotemark[$\dagger$] & 0.8$\pm$0.2      & 1.9$\pm$0.3                         & 2.5$\pm$0.6                        & 1.1$\pm$0.3             & 0.7$\pm$0.3               \\
		\hline
		$\chi^2$/d.o.f.  &                                               & 324.00/285                          & 250.48/199                           & 116.43/122                           & 144.97/130              & 92.73/77                     \\ 
		\hline \vspace{-3mm} \\
	\end{tabular}
	}
	\label{tab:ctb1ABCDE_best_fit_parameter}
	\begin{tabnote}
		\hangindent6pt\noindent
		\hbox to6pt{\footnotemark[$*$]\hss}\unskip%
		Normalization of the APEC component is $\frac{1}{4\pi D^2} \int n_en_\mathrm{H}dV$, where $D$ is the distance to the source, $n_e$ and $n_\mathrm{H}$ are the electron and hydrogen densities, respectively, and $V$ is the X-ray emitting volume.
		\par
		\hangindent6pt\noindent
		\hbox to6pt{\footnotemark[$\dagger$]\hss}\unskip%
		Unit of 10$^{-15}$ erg~cm$^{-2}$~s$^{-1}$~arcmin$^{-2}$ in the 2.0--10.0~keV band.
	\end{tabnote}
\end{table*}

\section{Discussion}
\subsection{Absorption column density toward CTB~1}
From the spatially resolved analysis (section \ref{secSRA}), we obtained the absorption column densities of the five regions.
The column densities of the inner radio-shell regions (regions C, D and E) are about 2 times higher than those of the breakout regions (regions A and B).
This result might indicate that there is a local enhancement of the ISM density around the radio shell of CTB~1.
Indeed, the interaction between the CTB~1 shell and a surrounding H\emissiontype{I} gas was reported by \citet{YarUyaniker2004}. 
However, excess of a hydrogen column density estimated from the H\emissiontype{I} gas interacting with CTB~1 ($T_\mathrm{B} \sim 30$~K at $v_\mathrm{LSR} = -15.5$--$-18.5$~km/s) is $1.6\times10^{20}$~cm$^{-2}$, which is one order of magnitude smaller than that of our observations.
The survey of $^{12}$CO ($J$ = 1--0) in this area was performed by \citet{heyer1998}, but no association between CTB~1 and molecular clouds have been reported so far.
Further observations of the cold ISM are necessary to identify the origin of the excess of the X-ray absorption column density at the shell of CTB~1.

We estimated the distance to CTB~1 according to the absorption column density toward the breakout region ($N_\mathrm{H} \sim 2.3\times10^{21}$~cm$^{-2}$).
Assuming the mean interstellar hydrogen density of 1~cm$^{-3}$, the distance is to be $\sim 0.8$~kpc.
This is the lowest value among estimation from other studies (1.6--3.1~kpc: \cite{craig1997}; \cite{YarUyaniker2004}).
However, the mean interstellar hydrogen density toward the CTB1 direction is highly uncertain.
If we adopt a hydrogen density of 0.4~cm$^{-3}$, we obtained a distance of 2~kpc, which is an intermediate value of the previous estimates. 
In the following discussion,  we assumed a distance of 2~kpc.
We confirmed that our conclusions are not affected even when the distance is changed within 0.8--3.1~kpc.

\subsection{Recombining plasma in the SW region}
Previous observations with Chandra and ASCA suggested that both the NE and SW plasmas of CTB~1 are in the CIE state with $kT_e$ of 0.2--0.3~keV (\cite{lazendic2006}; \cite{pannuti2010}).
However, our spectral analyses revealed that the SW plasma is not in CIE but in recombining phase for the first time.

Previous studies of RPs indicate that it is effective for the investigation of the formation process of the RPs to compare between the distributions of densities of ISM and $kT_e$ of the RPs (W49B: \cite{lopez2013}; IC~443: \cite{matsumura2017b}; W28: \cite{okon2018}).
\citet{lopez2013} discussed that higher $kT_e$ would be expected in a region with a high ambient gas density if they consider the rarefaction as the formation process of RPs.
On the contrary, \citet{matsumura2017b} and \citet{okon2018} found that $kT_e$ of RPs in shock-cloud interaction regions is significantly lower than those in the other region, suggesting that the origin of the RPs is cooling by thermal conduction between the SNR plasma and the cool dense ISM.

\begin{figure}[t]
	\begin{center}
	\includegraphics[width=80mm]{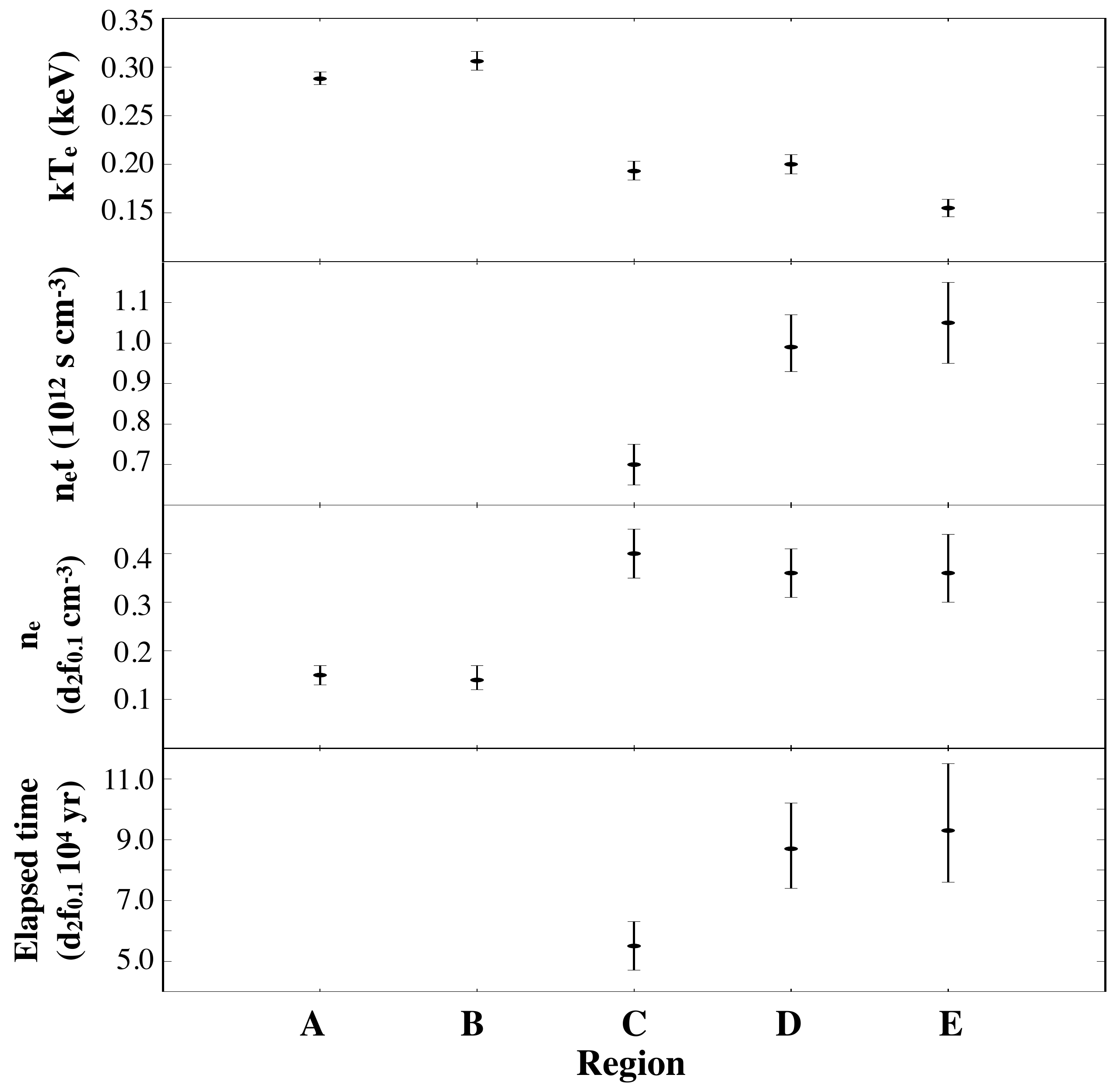}
	\end{center}
	\caption{Top panel shows electron temperatures in the regions A--E with a unit of keV. 
	Secand panel shows ionization parameters in the regions A--E with a unit of 10$^{12}$~s~cm$^{-3}$.
	Third panel shows the densities of electrons.
	$d_{2}$ is the distance to CTB~1 divided by 2~kpc and $f_{0.1}$ is the filling factor divided by 0.1.
	The lower panel shows elapsed times for each region.
	}
	\label{fig:ctb1_pars_nei}
\end{figure}

CTB~1 provides a unique opportunity to test the formation process of the RP; 
its ambient density is low at the breakout region and high at the shell region \citep{YarUyaniker2004}. 
As shown in figure~\ref{fig:ctb1_pars_nei}, we found that $kT_e$ in the inner-shell region is lower than that in the breakout region.
Moreover, the rim of the shell (region E) shows the lowest $kT_e$.
Therefore, the RP in CTB~1 is most likely explained by the thermal conduction scenario rather than the rarefaction scenario. 

In addition, we calculated the elapsed time since recombination started to dominate over ionization ($t_\mathrm{rec}$) using $n_et$ obtained from our spectral analyses.
The bottom two panels of figure~\ref{fig:ctb1_pars_nei} show the electron densities ($n_e$) estimated from the best-fit emission measures and the derived $t_\mathrm{rec}$ for the region C, D, and E.
The result suggests that the outer region became an RP earlier than the central region did.
It is consistent with the thermal conduction scenario.

\subsection{Spatial variation of the metal abundances}
The abundances of Ne and Mg are larger than the solar values in both the NE and SW regions and are consistent with results of previous studies (\cite{lazendic2006}; \cite{pannuti2010}).
In contrast, we discover that the abundance of Fe in the SW region is much higher than that in the NE region.
We note that the abundance of Fe is significantly affected by an assumed model as shown in a comparison between the SW Model1 and the SW Model2 (see table \ref{tab:ctb1sw_best_fit_parameter}), and therefore precise modeling of the plasma ionization state is crucial to determine the metal abundances.

\begin{figure}[t]
	\begin{center}
	\includegraphics[width=80mm]{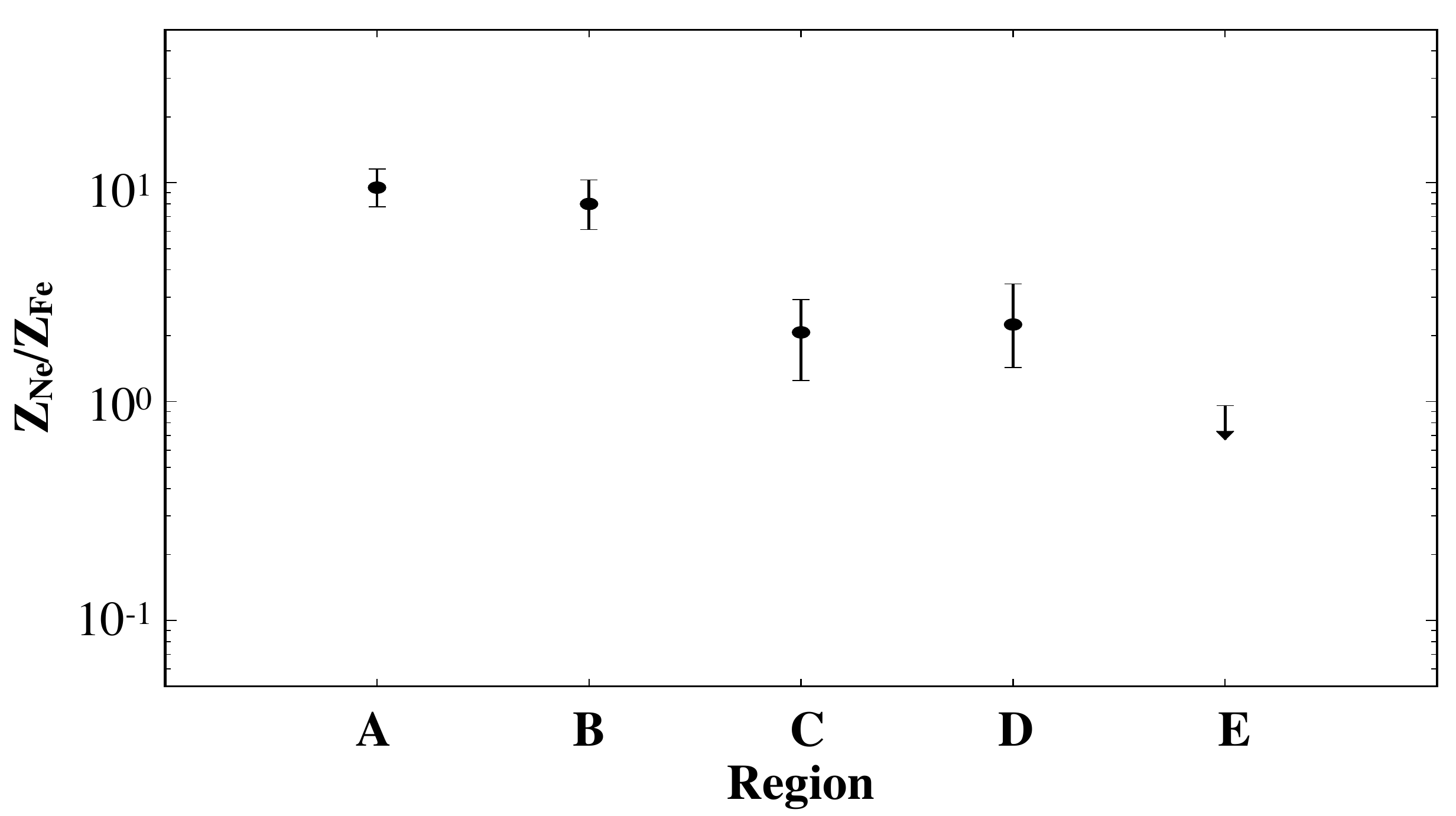}
	\end{center}
	\caption{
	The abundance ratios of Ne to Fe in the regions A--E. 
	}
	\label{fig:ctb1_pars_abd}
\end{figure}

In figure \ref{fig:ctb1_pars_abd}, We plot a spatial variation in the abundance ratio of Ne to Fe.
The ratios are systematically larger than unity suggesting the origin of a core-collapse supernova (e.g., \cite{nomoto2006}; \cite{takeuchi2016}).
In addition, the ratios in the inner-radio shell region are smaller than those in the breakout region.
In particular, Fe is enhanced in the region E.
This indicates an asymmetric ejecta distribution as also found in other SNRs with the origin of core-collapse supernovae (e.g., \cite{katsuda2008}; \cite{uchida2009}).
One of the possibilities is that Ne was isotropically ejected while Fe was ejected toward SW.

\subsection{Origin of the hard-band excess}
\begin{figure}[t]
	\begin{center}
	\includegraphics[width=80mm]{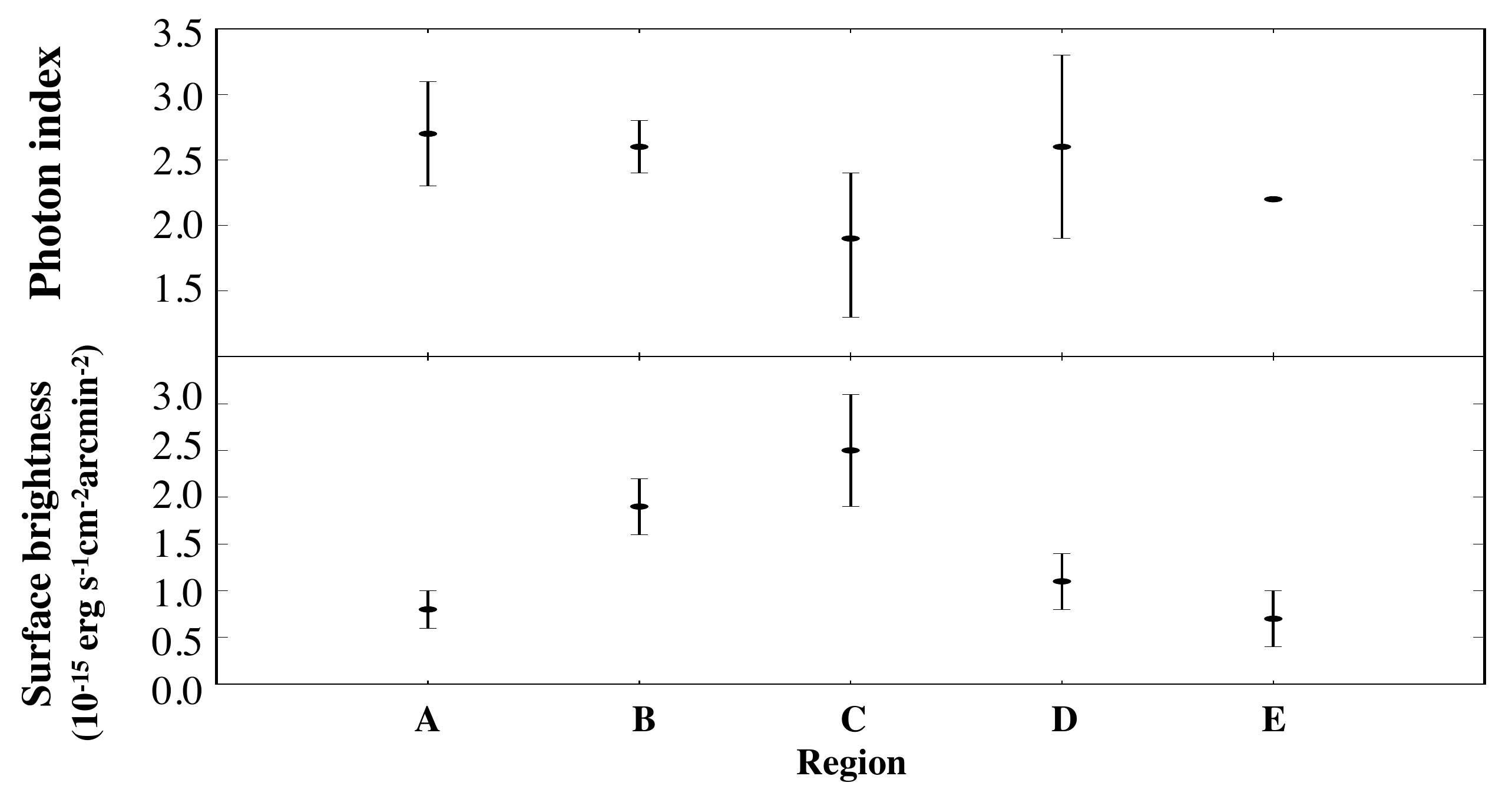}
	\end{center}
	\caption{
	Top panel shows the photon index in the regions A--E. 
	The photon index for the region E is fixed at the best-fit value of the entire SW region (2.2).
	The lower panel shows the surface brightness (2.0--10~keV) in the regions A--E with a unit of 10$^{-15}$~erg~cm$^{-2}$s$^{-1}$arcmin$^{-2}$.
	}
	\label{fig:ctb1_pow_pars}
\end{figure}

We found that the observed fluxes above 2~keV are larger than that of the CXB model by 
$\sim$13--47\% in both NE and SW.
These hard-band excesses are represented by power-law functions.
Photon indexes and surface brightnesses for all the regions are shown in figure  \ref{fig:ctb1_pow_pars}.
$\Gamma$ is in the range of 2--3, and marginal spectral steeping is seen at the outer of CTB~1. 
On the other hand, the surface brightness has a peak at the center of the radio shell.
A flux and a luminosity integrated over the observed regions are $4.5\times10^{-13}$~erg~s$^{-1}$cm$^{-2}$ and $2.2\times10^{32}d_{2}^2$~erg~s$^{-1}$ in the 2--10~keV band, respectively, where $d_{2}$ is the distance to CTB~1 divided by 2~kpc.
A spatial extent of the hard X-ray emission is 20 $d_{2}$~pc.

 \citet{pannuti2010} reported the detection of power-law components in the NE and SW regions using ASCA spectra.
The photon indexes and the fluxes they derived are consistent with those of our results.
They also argued that a thermal plasma with $kT_e \sim 3$~keV, which might be a high-temperature plasma component in CTB~1, can explain the observed spectra instead of the power-law component.
However, fitting a thermal plasma model to the Suzaku spectra (NE Model2) requires an extremely low abundance due to a lack of emission lines from He-like Fe.
Because such a low abundance is not explained by either a supernova ejecta nor the ISM, the high-temperature plasma associated with CTB~1 is unlikely.

One possible origin of the hard-band excess is a flux fluctuation of the CXB due to the cosmic variance.
Indeed, an expected amplitude of the fluctuation for the Suzaku FoV is $\sim 15$\% \citep{moretti2009}.
In order to confirm the CXB fluctuation in the two background regions, we fitted the background spectra again with the free CXB normalizations, and found that the difference in the CXB normalizations between BG1 and BG2 is 14\%, which is consistent with the above expectation.
On the other hand, the hard-band excesses in CTB1 are 13--47\% of the CXB flux and are higher than the expected fluctuation in the regions B, C, and D.
Moreover, the derived photon index is significantly steeper than that of the CXB, and the excesses are observed in all regions.
Therefore, the origin of the CXB fluctuation is unlikely.

On the basis of the morphology, no clear association is seen between the X-ray hard-band excess and the radio shell.
One possible origin is a pulsar wind nebula, because $\Gamma$, the luminosity, and the spatial extent is in the range of old Galactic pulsar wind nebulae (\cite{kargaltsev2008};  \cite{bamba2010}). 
According to the empirical relation between the luminosity and the characteristic age \citep{mattana2009}, the characteristic age is to be 10$^4$--10$^5$ yr, which is consistent with the age of CTB~1.
The spin down luminosity of the pulsar is estimated to be $2.8\times10^{35}$~erg~s$^{-1}$ assuming $L_\mathrm{x}/\dot{E}_\mathrm{rot} = 8\times10^{-5}$ for old pulsar wind nebulae \citep{vink2011}.
However, no pulsar has been detected at the center of CTB~1.
Moreover, gamma-ray emission from a pulsar wind nebula has not been detected so far even though an expected TeV gamma-ray luminosity is $\sim10^2 L_\mathrm{x}$ according to the relation shown by  \citet{mattana2009}.  
Further multi-wavelength observations are necessary to examine this scenario.

\section{Summary}
We observed  MM SNR CTB~1 with Suzaku for the total exposure of $\sim 82$~ks, and obtained the following results from the spectral analysis.
\begin{enumerate}\raggedright
	\item The 0.6--2.0~keV spectra in the NE breakout region of CTB~1 were fitted with a CIE plasma model with $kT_e \sim 0.3$~keV, 
	whereas those in the SW inner-shell region were well represented by an RP model with $kT_e  \sim 0.19$~keV, $kT_\mathrm{init} = 3.0$~keV and $n_et \sim 9\times10^{11}$~cm$^{-3}$s.
	This is the first detection of an RP in CTB~1.
	\item The characteristic morphology of CTB~1 provides unique opportunity to test formation scenarios of RPs.
	Our results show that kTe in the inner-shell region is lower than that in the breakout region, and becomes lowest at the rim of the shell.
	In addition, $t_{rec}$ increases toward the outer region.
	Therefore, the thermal conduction scenario is likely for the formation of the RP in CTB~1 rather than the rarefaction scenario.
	\item The Ne abundance is almost uniform in the observed regions, whereas the Fe abundance is enhanced in the inner-shell region, suggesting the asymmetric ejecta distribution.
	\item The diffuse hard X-ray emission represented by a power-law function is detected. 
	The photon index is $\sim 2.5$ and the total flux is $\sim 5 \times 10^{-13}$ erg cm$^{-2}$ s$^{-1}$ in the 2--10~keV band.
	The surface brightness is peaked at the center of CTB~1.
	The origin of this emission is an open question but one possibility is a pulsar wind nebula associated with this remnant.
  \end{enumerate}

\begin{ack}
We appreciate all the Suzaku team members. 
This work is supported by Japan Society for the Promotion of Science (JSPS) KAKENHI grant number 16J02332 (M. K.), 18J01417 (H. M.), 16H02170(T. T.) and World Premier International Research Center Initiative (WPI), MEXT, Japan.
S. N. is supported by the RIKEN Special Postdoctoral Researcher Program.
SHL is supported by The Kyoto University Foundation (grant no. 203180500017).
M. A. is supported by RIKEN Junior Research Associate Program.
\end{ack}


\end{document}